\documentclass[prd,superscriptaddress,nofootinbib,amsmath,amssymb,aps,11pt]{revtex4-2}

\usepackage{bm}
\usepackage{amsfonts}
\usepackage{latexsym}
\usepackage[latin1]{inputenc}
\usepackage{graphicx}
\usepackage{amsmath}
\usepackage{palatino}
\usepackage{mathpazo}
\usepackage{booktabs}
\usepackage{dcolumn}

\def\jnl@style{\it}
\def\aaref@jnl#1{{\jnl@style#1}}

\def\aaref@jnl#1{{\jnl@style#1}}

\def\aj{\aaref@jnl{AJ}} 
\def\apj{\aaref@jnl{ApJ}} 
\def\apjl{\aaref@jnl{ApJ}} 
\def\apjs{\aaref@jnl{ApJS}} 
\def\apss{\aaref@jnl{Ap\&SS}} 
\def\aap{\aaref@jnl{A\&A}} 
\def\aapr{\aaref@jnl{A\&A~Rev.}} 
\def\aaps{\aaref@jnl{A\&AS}} 
\def\mnras{\aaref@jnl{Mon.~Not.~Roy.~Astron.~Soc.}} 
\def\prd{\aaref@jnl{Phys.~Rev.~D}} 
\def\prc{\aaref@jnl{Phys.~Rev.~C}} 
\def\prl{\aaref@jnl{Phys.~Rev.~Lett.}} 
\def\qjras{\aaref@jnl{QJRAS}} 
\def\skytel{\aaref@jnl{S\&T}} 
\def\ssr{\aaref@jnl{Space~Sci.~Rev.}} 
\def\zap{\aaref@jnl{ZAp}} 
\def\nat{\aaref@jnl{Nature}} 
\def\aplett{\aaref@jnl{Astrophys.~Lett.}} 
\def\apspr{\aaref@jnl{Astrophys.~Space~Phys.~Res.}} 
\def\physrep{\aaref@jnl{Phys.~Rep.}} 
\def\physscr{\aaref@jnl{Phys.~Scr}} 
\def\commat{\aaref@jnl{Comm.~Math.~Phys.}} 
\def\science{\aaref@jnl{Science}} 
\def\cqg{\aaref@jnl{Classical Quant.~Grav.}} 
\def\jpcs{\aaref@jnl{JPCS}} 
\def\ijmpd{\aaref@jnl{Int.~J.~Mod.~Phys.~D}} 
\def\grg{\aaref@jnl{Gen.~Relat.~Gravit.}} 
\def\rpp{\aaref@jnl{Rep.~Prog.~Phys.}} 
\def\npa{\aaref@jnl{Nucl.~Phys.~A}} 
\def\lrr{\aaref@jnl{Living Rev.~Rel.}} 
\def\jcap{\aaref@jnl{J.~Cosmology Astropart.~Phys.}} 
\def\rmp{\aaref@jnl{Rev.~Mod.~Phys.}} 

\allowdisplaybreaks[1]

\begin{document}

	\title{Black hole no-hair theorem for self-gravitating time-dependent spherically symmetric multiple scalar fields }
	
	\author{Stoytcho S. Yazadjiev}
	\email{yazad@phys.uni-sofia.bg}
	\affiliation{Department of Theoretical Physics, Faculty of Physics, Sofia University, Sofia 1164, Bulgaria}
	\affiliation{Institute of Mathematics and Informatics, 	Bulgarian Academy of Sciences, 	Acad. G. Bonchev St. 8, Sofia 1113, Bulgaria}

	\author{Daniela D. Doneva}
	\email{daniela.doneva@uni-tuebingen.de}
	\affiliation{Theoretical Astrophysics, Eberhard Karls University of T\"ubingen, T\"ubingen 72076, Germany}
	\affiliation{INRNE - Bulgarian Academy of Sciences, 1784 Sofia, Bulgaria}


	\begin{abstract}
		We prove under certain weak assumptions a black hole no-hair theorem in spherically symmetric spacetimes for self-gravitating time-dependent multiple scalar fields with an arbitrary target space admitting a Killing field with a non-empty axis and arbitrary non-negative potential invariant under the flow of the Killing field. It is shown that for such configurations the only spherically symmetric and asymptotically flat black hole solutions consist of the Schwarzschild metric and a constant multi-scalar map. In due course of the proof we also unveil the intrinsic connection of the time-dependence of the scalar fields with the symmetries of the target space. 
	\end{abstract}
	
	\maketitle
	
	\section{Introduction}
	
	Scalar fields play an important role in modern theoretical physics, especially in the context of gravitational theories, cosmology, and astrophysics. Of special interest for black hole physics is the question of whether black holes can support scalar hair. In the case of a single scalar field it is well known that black holes can not support scalar hair in physically realistic scenarios \cite{Bekenstein_1972},\cite{Herdeiro_2015}. The same applies to the case with multiple static scalar fields \cite{Heusler_1992},\cite{Heusler_1995}. The picture changes drastically in the case of time-dependent multiple scalar fields. Even for Einstein's equations minimally coupled to two (real) scalar fields or equivalently one complex scalar field, stationary black hole solutions with scalar hair exist when the scalar fields are time-dependent but with a stationary energy-momentum tensor \cite{Herdeiro_2014},\cite{Collodel_2020a}. There are serious numerical indications that the picture of the possible stationary black hole solutions in the presence of two real time-dependent scalar fields (one time-dependent complex scalar field) could be very complicated \cite{Herdeiro_2023}. This fact naturally raises the problem for the classification of the possible stationary black hole solutions of the Einstein equations coupled to time-dependent multiple scalar fields. In its full generality, this mathematical problem is extremely difficult and that is why it is natural as a first step to focus on static black hole solutions with spherical symmetry. A partial result, in this case, has been achieved in \cite{Pena_1997} where a no-hair theorem for spherically symmetric scalar fields with special time dependence in the particular case of a flat target space was proven. 
	
	The purpose of the present paper is to prove a black hole non-hair theorem in spherically symmetric spacetimes for self-gravitating time-dependent multiple scalar fields with an arbitrary target space and an arbitrary non-negative potential. The only non-avoidable restriction we impose on the target space metric is to admit a Killing field with a non-empty axis whose action leaves the potential invariant. This restriction, as explained below, ensures the stationarity of the energy-momentum tensor of the scalar fields. In due course of the proof we also unveil the intrinsic connection of the time-dependence of the scalar fields with the symmetries of the target space.

	\section{Black hole no-hair theorem for self-gravitating time-dependent spherically symmetric multiple scalar fields } 
	
	We consider the 4-dimensional spacetime $(M, g_{\mu\nu})$, an $N$-dimensional Riemannian manifold $({\cal E}_N,\gamma_{ab})$, the so-called target space, and a map $\varphi: (M,g_{\mu\nu})\to ({\cal E}_N,\gamma_{ab})$. The differential $d\varphi$ induces a map between the tangent spaces of $M$ and ${\cal E}_N$, $d\varphi: TM\to T{\cal E}_N$. The norm of the differential $<d\varphi,d\varphi>$ in local coordinate patches on $M$ and ${\cal E}_N$ is given by 
	\begin{eqnarray}
		<d\varphi,d\varphi>= g^{\mu\nu}(x)\gamma_{ab}(\varphi(x))\partial_\mu\varphi^a(x)\partial_\nu\varphi^b(x).
	\end{eqnarray} 
	
	We shall also consider multi-scalar self-gravitating theories defined by the following action
	\begin{eqnarray}\label{action}
		S=\frac{1}{16\pi G} \int_{M} d^4x\sqrt{-g}\left[R - 2<d\varphi,d\varphi> - 4V(\varphi)\right],
	\end{eqnarray}	
	where $R$ is the Ricci scalar curvature and $V(\varphi)$ is the scalar fields potential. The field equations associated with the action (\ref{action}) and written in local coordinate patches of $M$ and ${\cal E}_N$ are 
	\begin{eqnarray}\label{FE1}
		&&R_{\mu\nu} - \frac{1}{2}R g_{\mu\nu} = 2\gamma_{ab}(\varphi)\nabla_{\mu}\varphi^a \nabla_{\nu}\varphi^b - 
		\gamma_{ab}(\varphi)\nabla_{\sigma}\varphi^a \nabla^{\sigma}\varphi^b g_{\mu\nu} - 2V(\varphi) g_{\mu\nu}, \\
		&&\nabla_{\mu}\nabla^{\mu}\varphi^a= 
		- \gamma^{a}_{cd}(\varphi)\nabla_{\mu}\varphi^c \nabla^{\mu}\varphi^d + \frac{1}{4}\gamma^{ab}(\varphi)\frac{\partial V(\varphi)}{\partial\varphi^b}, \label{FE2}
	\end{eqnarray}
	where $\nabla_{\mu}$ is the covariant derivative with respect to the spacetime metric $g_{\mu\nu}$ and 
	$\gamma^{a}_{cd}(\varphi)$ are the Christoffel symbols with respect to the target space metric $\gamma_{ab}(\varphi)$. We shall assume that the potential $V(\varphi)$ is non-negative and that the system $\frac{\partial V(\varphi)}{\partial\varphi^a}=0$ possesses at least one solution $\varphi_{0}$ at which
	$V(\varphi_{0})=0$.

	We focus here on static and spherically symmetric spacetimes with a metric 
	\begin{eqnarray}
		ds^2= - e^{2\Phi(r)}dt^2 + e^{2\Lambda(r)}dr^2 + r^2 s_{ij}dx^idx^j,
	\end{eqnarray} 
	where $s_{ij}$ is the metric on the unit 2D sphere, namely $s_{ij}dx^idx^j=d\theta^2 + \sin^2\theta d\phi^2$.
	
	We require the scalar fields to be spherically symmetric, i.e. invariant under the flow of the Killing fields generating the spherical symmetry. In simple words this means that the scalar fields are independent from the angular coordinate, i.e. $\partial_i\varphi^a=0$. Contrary to that, the scalar fields are not required to depend only on the radial coordinate $r$. They can depend on the time in such a way that the energy-momentum tensor to be time-independent. In order to guarantee that the energy-momentum tensor of the scalar fields is stationary (time-independent), we assume that the target space metric $\gamma_{ab}(\varphi)$ admits a Killing field $k^a$ whose flow leaves the potential $V(\varphi)$ invariant, namely ${\cal L}_{k}V(\varphi)=k^a\partial_a V(\varphi)=0$ and the dynamics of scalar fields is confined on the flow of $k^a$. In more formal language we require the Lie derivative ${\cal L}_{\xi} \varphi^a$ along the timelike Killing field $\xi=\frac{\partial}{\partial t}$ satisfies ${\cal L}_{\xi} \varphi^a=\partial_{t}\varphi^a = -\omega k^a$ where $\omega \ne 0$ is a real number. Indeed, using that ${\cal L}_{\xi} \varphi^a= -\omega k^a$ we have 
	\begin{eqnarray}
		{\cal L}_{\xi} T_{\mu\nu}= -\omega \left({\cal L}_{k}\gamma_{ab}(\varphi)\right) \left[2\nabla_{\mu}\varphi^a \nabla_{\nu}\varphi^b - \nabla_{\sigma}\varphi^a \nabla^{\sigma}\varphi^b g_{\mu\nu}\right] + 2\omega {\cal L}_{k} V(\varphi)g_{\mu\nu} =0 .
	\end{eqnarray} 
	
	In order to further guarantee the staticity of the metric we must impose the condition $\xi^\sigma R_{\sigma[\mu}\xi_{\nu]}=0$, which, taking into account that $\partial_i\varphi^a=0$, reduces to $k_a \partial_{r}\varphi^a =0$. The norm of $k^a$ will be denoted by $|k|$, i.e. $|k|^2=\gamma_{ab}(\varphi)k^ak^b$. 
	
	From now on we assume the existence of such a Killing field $k^a$ and in addition we assume that the axis of $k^a$, i.e. the points where $k^a=0$, is non-empty. 
	The additional requirement for the axis of $k^a$ is important because of the following. First, as we discuss below, the asymptotic flatness requires $\lim_{r\to \infty}|k|=0$. Second, in order for the black hole to be static on the black hole horizon we must have $R_{\mu\nu}\xi^\mu\xi^\nu|_h= T_{\mu\nu}\xi^\mu\xi^\nu|_h=0$ which, using the Einstein equations and ${\cal L}_{\xi} \varphi^a = -\omega k^a$, reduces to $R_{\mu\nu}\xi^\mu\xi^\nu=\omega^2 |k|^2|_h=0$. All this clearly shows that the axis of $k^a$ must be non-empty. 
	
	Another way to derive the conditions $k_a\partial_r\varphi^a=0$ and $|k||_h=0$ is as follows. When the target space metric admits a Killing field leaving the scalar potential invariant, one can show by using the field equations that there exists a conserved current $j^\mu$ satisfying $\nabla_{\mu}j^\mu=0$ and explicitly given by 
	\begin{eqnarray}
		j^\mu=k_a \nabla^{\mu}\varphi^a.
	\end{eqnarray} 
	
	Then the condition $k_a\partial_r\varphi^a=0$ means that the scalar flux in radial direction vanishes. The scalar flux through the horizon is 
	$\xi^\mu j_{\mu}=\xi^\mu k_a\nabla_{\mu}\varphi^a=-\omega |k|^2|_h$. Therefore the condition $\left. |k|\right|_h=0$ guarantees that there is no scalar flux through the horizon and the geometry can be static.

	The assumptions we made above are very natural from mathematical and physical points of view. In particular, if the orbits of the Killing field $k^a$ are periodic, then the time-dependence will also be periodic. This is the case for example with boson stars which constitute of two time-dependent scalar fields periodic in time, or equivalently one periodic in time complex scalar field. Let us demonstrate this on a simple but instructive example. Consider the flat target space $(R, \delta_{ab})$. It admits three Killing vectors $k_{(1)}=\frac{\partial}{\partial\varphi^{1}}$ , $k_{(2)}=\frac{\partial}{\partial\varphi^{2}}$ and $k_{(3)}=\varphi^{2}\frac{\partial}{\partial\varphi^{1}}-\varphi^{1} \frac{\partial}{\partial\varphi^{2}} $. Only the third Killing vector has periodic orbits and non-empty axis corresponding to $\varphi^1=\varphi^2=0$. In this case, the conserved current associated with the Killing vector $k_{(3)}$ is $j^\mu={k^{(3)}}_a\nabla^\mu \varphi^a=\varphi^2\nabla^\mu \varphi^1 - \varphi^2\nabla^\mu \varphi^1 = \frac{i}{2}(\Psi^*\nabla^\mu \Psi - \Psi\nabla^\mu \Psi^* )$
	which is the standard conserved current for the complex scalar field $\Psi=\varphi^1 + i\varphi^2$. Solving then the equations $\frac{\partial\varphi^a}{\partial t}= -\omega k^a_{(3)}$ and taking into account that $\varphi^a$ do not depend on the angular coordinates, we find 
	\begin{eqnarray}
		\varphi^{1}(r,t)=\psi(r)\cos(\omega t + f(r)), \, \, \ \varphi^{2}(r,t)=\psi(r)\sin(\omega t + f(r)),
	\end{eqnarray} 
	where $\psi(r)$ and $f(r)$ are functions of the radial coordinate only. Imposing further the staticity condition $k_a \partial_{r}\varphi^a =\varphi^{2}\partial_{r}\varphi^1 - \varphi^{1}\partial_{r}\varphi^2 =0$ we obtain $\psi^2(r)\partial_r f(r)=0$ which means that $f(r)=C=constant$.
	Without loss of generality we can put $C=0$. In terms of the complex scalar field $\Psi=\varphi^1 + i\varphi^2$ we have $\Psi(r,t)=\psi(r)e^{i\omega t}$. In this way we derived the standard ansatz for the complex scalar field in studying the boson stars and the black holes.

	Let us go back to the general case. We shall make one more technical assumption. In order to satisfy the asymptotic flatness we require the axis of $k^a$ to contain the solution $\varphi_0$ of the algebraic system $\frac{\partial V(\varphi)}{\partial\varphi^a}=0$. This can be easily achieved by assuming that the potential depends properly on the scalar field through the norm of the Killing filed $k^a$, $V(\varphi)=V(|k|^2)$, which is the common case with two (real) scalar fields with flat target space or equivalently one complex scalar field, considered in the literature\footnote{In the literature the potential $V(\varphi)$, expressed in terms of the norm of the complex scalar field $\Psi$, is usually in the form $V(\varphi)=2m^2|\Psi|^2 + \lambda |\psi|^4$ with $|\Psi|^2=(\varphi^1)^2 + (\varphi^2)^2=|k_{(3)}|^2$.}.
	
	Under the assumptions we made, the dimensionally reduced field equations are the following 
	\begin{eqnarray}
		&&\frac{2}{r}e^{-2\Lambda} \Lambda^{\prime} + \frac{1}{r^2}\left(1-e^{-2\Lambda}\right)= \omega^2 e^{-2\Phi}|k|^2 + 
		e^{-2\Lambda}\gamma_{ab}(\varphi)\partial_r\varphi^a\partial_r\varphi^b + 2V(\varphi), \label{DRE} \notag \\ \\ 
		&&\frac{2}{r}e^{-2\Lambda} \Phi^{\prime} - \frac{1}{r^2}\left(1-e^{-2\Lambda}\right)= 
		\omega^2 e^{-2\Phi}|k|^2 + 
		e^{-2\Lambda}\gamma_{ab}(\varphi)\partial_r\varphi^a\partial_r\varphi^b - 2V(\varphi), \notag \\ \\
		&&e^{-2\Lambda}\left[\Phi^{\prime\prime} + (\Phi^{\prime} + \frac{1}{r})(\Phi^{\prime} - \Lambda^{\prime})\right]=
		-\frac{1}{2}K - 2V(\varphi), \notag \\ \notag \\ \\
		&&\partial_r \left(e^{\Phi-\Lambda} r^2 \gamma_{ab}(\varphi)\partial_r\varphi^b \right) = 
		\frac{1}{4}e^{\Phi + \Lambda} r^2 \left(\frac{\partial V(\varphi)}{\partial \varphi^a} + \frac{\partial K}{\partial \varphi^a} \right) +\omega^2 r^2 e^{\Lambda -\Phi} k^b \frac{\partial k_a}{\partial \varphi^b},\label{EQFF2} \notag \\
	\end{eqnarray} 
	where $K$ is defined by 
	\begin{eqnarray}\label{K}
		K= 2\gamma_{ab}(\varphi)\nabla_{\mu}\varphi^a\nabla^{\mu}\varphi^b = -2\omega^2 e^{-2\Phi}|k|^2 + 
		2e^{-2\Lambda}\gamma_{ab}(\varphi)\partial_r\varphi^a\partial_r\varphi^b . \notag \\
	\end{eqnarray} 
	
	Taking into account that the metric functions $\Phi$ and $\Lambda$ depend on $r$ only, from the dimensionally reduced field equations it follows that $\gamma_{ab}(\varphi)\partial_r\varphi^a \partial_r\varphi^b $ and $|k|$ are functions of $r$ only. Then we can define
	\begin{eqnarray}
		P^2(r)=\gamma_{ab}(\varphi)\partial_r\varphi^a \partial_r\varphi^b .
	\end{eqnarray}
	
	In the present paper we are interested in asymptotically flat spacetimes. In this case, by using the dimensionally 
	reduced field equations, one can easily see that 
	\begin{eqnarray}\label{BCAINF}
		\lim_{r\to \infty} P^2(r)=0, \;\; \; \lim_{r\to \infty} V(\varphi)=0, \;\;\; \lim_{r\to \infty} |k|=0. 
	\end{eqnarray}
	On the regular horizons the function $(\Phi + \Lambda)$ and its derivative are regular. Using this fact, and by adding the equation for $\Phi$ to the equation for $\Lambda$,
	one can see that 
	\begin{eqnarray}
		\lim_{r\to r_h} e^{-2\Phi}|k|^2=0 . 
	\end{eqnarray}
	
	The strategy for proving the no-hair theorem is based on a proper divergence identity. In order to \textbf{derive it} we used the following methodology. We multiply the equation (\ref{EQFF2}) for $\varphi^a$ by $\partial_r\varphi^a$ and after some algebra we obtain 
	\begin{eqnarray}\label{I12}
		\partial_r \left(e^{\Phi-\Lambda} r^2 \gamma_{ad}(\varphi)\partial_r\varphi^a \partial_r\varphi^d \right) - 
		\left(e^{\Phi-\Lambda} r^2 \gamma_{ad}(\varphi) \partial_r\varphi^d \right) \partial^2_r\varphi^a \nonumber \\ = 
		\frac{1}{4}e^{\Phi + \Lambda} r^2 \left(\frac{\partial V(\varphi)}{\partial \varphi^a} + \frac{\partial K}{\partial \varphi^a} \right) \partial_r\varphi^a +\omega^2 r^2 e^{\Lambda -\Phi} k^b \frac{\partial k_a}{\partial \varphi^b} \partial_r\varphi^a . \notag
	\end{eqnarray}
	We then express $e^{\Phi-\Lambda} r^2 \partial_r^2 \varphi^a$
	again from (\ref{EQFF2}) and substitute into the above equation (\ref{I12}). Doing so we get
	\begin{eqnarray}\label{I22}
		\partial_r \left(e^{\Phi-\Lambda} r^2 \gamma_{ad}(\varphi)\partial_r\varphi^a \partial_r\varphi^d \right) + 
		\partial_r\left( r^2 e^{\Phi-\Lambda}\gamma_{ad}(\varphi)\right) \partial_r\varphi^a \partial_r\varphi^d \nonumber \\ = 
		\frac{1}{2}e^{\Phi + \Lambda} r^2 \left(\frac{\partial V(\varphi)}{\partial \varphi^a} + \frac{\partial K}{\partial \varphi^a} \right) \partial_r\varphi^a + 2\omega^2 r^2 e^{\Lambda -\Phi} k^b \frac{\partial k_a}{\partial \varphi^b} \partial_r\varphi^a . \notag 
	\end{eqnarray}
	The next step is to make use of the explicit form (\ref{K}) of $K$ and after long algebra we find 
	\begin{eqnarray}\label{I3i}
		&&\frac{d}{dr}\left[ 2 r^2 e^{\Phi+\Lambda}V(\varphi) - r^2 e^{\Phi-\Lambda} P^2 - \omega^2 r^2 e^{\Lambda-\Phi} |k|^2 \right] \\
		&&=r e^{\Phi + \Lambda} \left[4V(\varphi) + (1+ e^{-2\Lambda})P^2 + \omega^2 e^{-2\Phi} |k|^2 (e^{2\Lambda} -3) \right].\nonumber 
	\end{eqnarray}
	This divergence identity is however not what we want since the right hand side is not manifestly non-negative. Another divergence identity with the desired property can be derived in the following way. We write our identity (\ref{I3i}) in the form 
	\begin{eqnarray}
		\frac{dB}{dr}= A.
	\end{eqnarray}
	Then if $\Omega(r)$ is an everywhere positive function, we can construct a new identity of the form
	\begin{eqnarray}
		\frac{d (\Omega B)}{dr}= \Omega \left(A + \frac{\Omega^\prime}{\Omega}B\right). 
	\end{eqnarray} 
	Applying this to our identity (\ref{I3i}) with $\Omega(r)=r^{-2}$  we finally obtain the desired divergence identity, namely 
	\begin{eqnarray}\label{I32}
		&&\frac{d}{dr}\left[ 2 e^{\Phi+\Lambda}V(\varphi) - e^{\Phi-\Lambda} P^2 - \omega^2 e^{\Lambda-\Phi} |k|^2 \right] \\
		&&= \frac{1}{r}e^{\Phi + \Lambda} \left[(1+ 3e^{-2\Lambda})P^2 + \omega^2 e^{2\Lambda -2\Phi} |k|^2 (1- e^{-2\Lambda}) \right].\nonumber 
	\end{eqnarray}
	The right side of the above equation is non-negative which follows from the fact that $(1-e^{-2\Lambda})>0$ for $r\in [r_H,\infty)$ and this can be seen from 
	the equation for $\Lambda$ written in the form 
	\begin{eqnarray}
		\frac{d}{dr}\left[r(1-e^{-2\Lambda})\right] = r^2\left[\omega^2 e^{-2\Phi}|k|^2 + 
		e^{-2\Lambda}P^2(r) + 2V(\varphi)\right]\ge 0.\notag 
	\end{eqnarray}
	Therefore $r(1-e^{-2\Lambda})$ is non-decreasing function and taking into account that it is positive on the horizon and $r>0$, we conclude that $(1-e^{-2\Lambda})>0$ 
	for $r\in[r_{H},+\infty)$.
	
	The last step is to integrate the identity (\ref{I32}) from the horizon to infinity and we get 
	\begin{eqnarray}\label{I3}
		- 2e^{(\Phi + \Lambda)_{h}} V(\varphi_{h})	= \int^{+\infty}_{r_H} dr\frac{1}{r}e^{\Phi + \Lambda} \left[(1+ 3e^{-2\Lambda})P^2 + \omega^2 e^{2\Lambda -2\Phi} |k|^2 (1- e^{-2\Lambda}) \right],\nonumber 
	\end{eqnarray}
	where we have taken into account (\ref{BCAINF}) and that $\lim_{r\to r_h}e^{\Phi-\Lambda}P^2(r)=\lim_{r\to r_h}e^{\Lambda-\Phi}|k|^2=0$. The left hand side is non-positive since by assumption $V(\varphi)\ge 0$, while the right hand side is non-negative.
	Therefore we can conclude that both sides vanish. Consequently we have that $P^2(r)=\gamma_{ab}(\varphi)\partial_{r}\varphi^a\partial_{r}\varphi^b=0$ and $\omega^2|k|^2=\gamma_{ab}(\varphi)\partial_{t}\varphi^a\partial_{t}\varphi^b=0$ everywhere which means that the scalar fields are constant in the domain of outer communications. Therefore the right hand side of the dimensionally reduced field equations (\ref{DRE})--(\ref{EQFF2}) vanish and the equations are reduced to the vacuum, static and spherically symmetric Einstein equations whose unique black hole solution with regular horizon is the Schwarzschild one. 
	
	Summarizing we proved the following theorem
	\medskip
	\noindent
	
	{\bf Theorem } {\it Let us consider self-gravitating map $\varphi: (M,g_{\mu\nu}) \to ({\cal E}_{N}, \gamma_{ab}) $ with the action (\ref{action}) and assume that the scalar field potential is non-negative, $V(\varphi)\ge 0$, and the system $\frac{\partial V(\varphi)}{\partial\varphi^a}=0$ possesses at least one solution $\varphi_{0}$ at which $V(\varphi_{0})=0$. Assume further that the scalar fields are invariant under the flow of the Killing fields generating the spherical symmetry, the target space metric admits a Killing field $k^a$ with non-empty axis containing roots $\varphi_0$ of the algebraic system $\frac{\partial V(\varphi)}{\partial\varphi^a}=0$ and the time dependence of the scalar fields satisfies ${\cal L}_{\xi}\varphi^a=-\omega k^a$ with $\omega\ne 0$ being a real number. Then every static and spherically symmetric black hole solution to the field equations (\ref{FE1})-(\ref{FE2}) with regular horizon consists of the Schwarzschild solution and a constant map $\varphi_0$ with $\frac{\partial V}{\partial\varphi^a}(\varphi_0)=V(\varphi_0)=|k|(\varphi_0)=0$. } 
	\medskip
	\noindent
	
	\section{Discussion}
	
	In the present paper we studied the black hole solutions to the Einstein gravity coupled to time-dependent multiple scalar fields with an arbitrary target space and non-negative potential in spherical symmetry. We first showed the intrinsic connection of the symmetries of the target space with the time-dependence of the scalar fields which ensures the stationarity of the energy-momentum tensor of the scalar field. Using that connection we derived a proper divergence identity with the help of which we proved a no-hair theorem. In this way, we achieved a full classification of the black hole solutions in the considered sector of the theory. The natural question is whether such a classification is possible for more general sectors of the theory, namely the static sector without spherical symmetry and stationary sector.  It turns out that, under certain additional assumptions, the full classification is also possible for the mentioned sectors of the theory. They require different mathematical techniques and will be presented elsewhere.

	\section*{Acknowledgments}
	This study is financed by the European Union-NextGenerationEU, through the National Recovery and Resilience Plan of the Republic of Bulgaria, project No. BG-RRP-2.004-0008-C01. DD acknowledges financial support via an Emmy Noether Research Group funded by the German Research Foundation (DFG) under grant
	no. DO 1771/1-1. 
	

\end{document}